# Hysteresis measurement of anomalous microwave surface resistance in superconducting thin films

X.S.Rao, C.K.Ong, B.B.Jin and Y.P.Feng

Centre for superconducting and Magnetic Materials and Department of Physics, National University of Singapore, Singapore 119260

The anomalous decrease in microwave surface resistance, $R_s$, of superconducting $YBa_2Cu_3O_{7-d}$ (YBCO) thin films in the presence of a low dc magnetic field is studied using a microstrip resonator technique. We have done a dc field hysteresis measurement of $R_s$ to study the effects of vortex penetration on the anomalous effect. It is shown that the anomaly happens at a field level far below the low critical field, $H_{c1,strip}$, of the superconducting microstrip and vortex (Abrikosov) penetration would eliminate the anomalous effect observed at low field. This implies that the anomalous effect is not contributed by vortices.

Recently, several low-field microwave surface resistance ($R_s$) anomalies, i.e., $R_s$ decreases at low dc magnetic field ($H_{dc}$) and microwave magnetic field ($H_{mw}$), has been observed [1-5]. In a previous paper [5], we studied the anomaly observed in YBCO thin films and showed that the $R_s(H_{dc})$ curve for $H_{dc}//c$-axis and $H_{dc}\perp c$-axis can coincide with each other through a normalization procedure and the anomaly was qualitatively independent of the dc field alignment. It is usually accepted that low-field microwave responses of HTS are mainly related with weak links and vortices [6,7]. As shown in [6], vortex mechanisms would have a pronounced angular dependence of the field effect on $R_s$, which comes from the anisotropy of YBCO. On the other hand, the angular dependence of the dc-field effect on $R_s$ for the weak-link mechanism is negligible since weak links are chaotically oriented [8]. Thus we suppose that the anomaly is not related to vortices and may come from weak links. To further confirm that vortices have no contribution on the anomaly observed, we have done a large dc field hysteresis measurement of the $R_s$. We have shown that the anomaly happens at a field level far below the lower critical field for vortex generation and vortices cannot account for the anomalous $R_s$ decrease.

The sample preparation and measurement were shown in [5]. The dc field is applied perpendicular with the film surface (//c-axis) using a Helmholtz coil configuration. The sample is initially cooled down to 77 K in zero field (ZFC). The dc field is ramped from zero to 2000 Oe and then decreased back to zero. The experimental results are shown in Fig.1 for the microstrip resonator working at 1.1 GHz with the input microwave power $P_{mw}$=-10 dBm.

The $R_s(H_{dc})$ curve in increasing field is characterized by a negative slope in the lowest field range ($H_{dc}$<6 Oe) where $R_s$ drops sharply as the dc field is increased. Then $R_s$ increases gradually with elevated $H_{dc}$. After that, a plateau, where the $R_s$ has a weaker dependence on field, is reached. This region extends roughly from $R_s(H_{dc})$~20 Oe to 70 Oe. At higher fields ($H_{dc}$>70 Oe) $R_s$ increases again and the slope of $R_s(H_{dc})$ curve is steep. After reaching the largest dc field, $H_{dc}^{max}$=2000 Oe, the dc field was subsequently decreased. In the field range $H_{dc}$>70 Oe the $R_s(H_{dc})$ curve in decreasing field coincides with that in increasing field and no hysteretic effect is found within the experimental error. When the dc field falls below 70 Oe, an obvious hysteretic loop can be observed in the figure where the $R_s(H_{dc})$ curve in decreasing field strongly deviates from that in increasing field. The $R_s(H_{dc})$ is almost constant in the region from 70 Oe to zero field. The final $R_s$ value at zero dc field is slightly larger than its initial value.

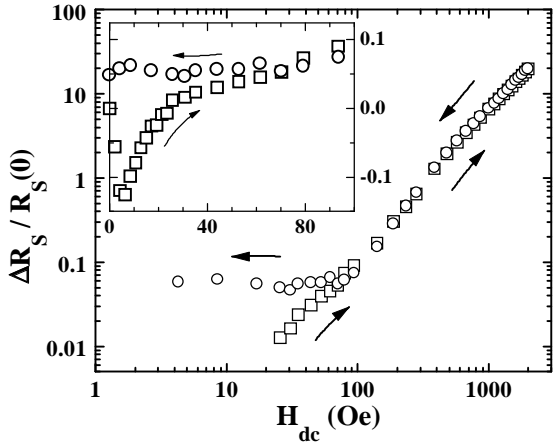

Figure 1. Hysteretic response of $R_s(H_{dc})$ for $H_{dc}^{max}$=2000 Oe.

The lower critical field of the superconducting YBCO strip, $H_{c1,strip}$~70 Oe, can be deduced from Fig.1. In the increasing field range, $R_s(H_{dc})$ is slightly dependent on the field below $H_{c1,strip}$ except the anomalous region (ranging from 0 to 20 Oe). At $H_{c1,strip}$, dc vortices (Abrikosov) begin to penetrate into the films and cause the steep increase in $R_s(H_{dc})$. When the field is increased up to 2000 Oe and then decreased below $H_{c1,strip}$, some vortices are trapped in the films due to pinning effect. This is the cause for the plateau of the $R_s(H_{dc})$ curve in decreasing field observed below 70 Oe. It is noticed that the $H_{c1,strip}$ (~70 Oe) is different from the bulk $H_{c1}$ for single crystalline YBCO measured by Wu *et al.* [9]. The difference may come from the large demagnetization effect [10] and the screening effect due to the superconducting ground plane. So it is clear that the anomalous decrease of $R_s$ happens at a field level (~6 Oe) far below the low critical field of the superconducting microstrip, where no vortex (Abrikosov) is formed. This is consistent with our previous result that shows the anomalous effect is qualitatively independent of the dc field alignment at low field [5].

The other important feature revealed by Fig.1 is that the anomalous field dependence of $R_s$ in weak dc field can only be observed in increasing field. In the $R_s(H_{dc})$ curve of decreasing field, the pinned vortices totally eliminate the anomalous effect observed in increasing field.

From the above results, we can conclude that the anomalous decrease of $R_s$ at low dc magnetic field is not contributed by vortices. So one may expect that the anomalous effect is related with weak links in the YBCO films. In fact, we have carried out a comparison study of two films with different film qualities and our preliminary results show that the film with more weak links has a stronger anomalous effect. A systematic study of the correlation between the anomalous effect and the microstructure of the YBCO thin films is in progress.